\documentclass[aps,prc,superscriptaddress,floats,floatfix,reprint,reprintnumbers]{revtex4}
\usepackage{blindtext}
\pdfoutput=1
\usepackage{epsfig}
\usepackage{graphicx}
\usepackage{dcolumn}
\usepackage{bm}
\usepackage{booktabs}
\usepackage{color}
\usepackage{mathrsfs}
\usepackage{ulem}
\usepackage{epstopdf}
\graphicspath{{plots/}}
\usepackage{verbatim}
\usepackage{indentfirst}
\usepackage{float}
\usepackage[colorlinks,
           bookmarks=true,
           linkcolor=blue,
           urlcolor=blue,
            anchorcolor=black,
            citecolor=blue]{hyperref}
\usepackage[bf,raggedright,indentafter]{titlesec}
\usepackage{rotating}
\usepackage{booktabs}
\usepackage{tabularx}
\usepackage{makecell}
\usepackage{hypcap}
\usepackage{subfigure}%
\usepackage{tabularx}

\usepackage{array}

\usepackage{amsmath}

\begin{document}

\title{\Large Study of semileptonic $B\to DP\ell^+\nu_\ell$  decays based on  the SU(3) flavor symmetry }

\author{Ru-Min Wang$^{1,\dagger}$,~~Yi-Jie Zhang$^{1}$,~~Meng-Yuan Wan$^{1}$,~~Xiao-Dong Cheng$^{2,\S}$,~~Yuan-Guo Xu$^{1,\sharp}$\\
 $^1${\scriptsize College of Physics and Communication Electronics, Jiangxi Normal University, Nanchang, Jiangxi 330022, China}\\
 $^2${\scriptsize College of Physics and Electronic Engineering, Xinyang Normal University, Xinyang, Henan 464000, China}\\
 $^\dagger${\scriptsize ruminwang@sina.com}~~
 $^\S${\scriptsize chengxd@mails.ccnu.edu.cn}~~
 $^\sharp${\scriptsize yuanguoxu@jxnu.edu.cn}~~
  }

\begin{abstract}
Decays $B\to DP\ell^+\nu_\ell~(\ell=e,\mu,\tau)$   with  the non-resonance, the charmed  vector resonances,  the charmed  scalar resonances and  the charmed  tensor resonances  are  explored
by using the SU(3) flavor symmetry approach.
Firstly, the decay amplitudes of  different modes are related by the SU(3) flavor symmetry. Then, relevant experimental data are used to constrain  non-perturbative coefficients in the
 non-resonant and various resonant $B\to DP\ell^+\nu_\ell$ decays. Finally, using the constrained non-perturbative coefficients, the branching ratios of not-yet-measured $B\to DP\ell^+\nu_\ell$  decays with
the non-resonant and various charmed  resonant contributions  are predicted. Many branching ratios are predicted for the first time.
We find that $B\to D\eta'\ell^+\nu_\ell, B_s\to D_s\eta'\ell^+\nu_\ell$ decays only receive the non-resonant contributions,   $B\to D_sK\ell^+\nu_\ell$, $B_s\to DK\ell^+\nu_\ell$, $B\to D\eta\ell^+\nu_\ell$ and
$B_s\to D_s\eta\ell^+\nu_\ell$ decays  receive both the non-resonant and the tensor resonant contributions,  $B^+\to D^-\pi^+\ell^+\nu_\ell$ decays receive the non-resonant, the  scalar resonant and the tensor resonant contributions, and other $B\to D\pi\ell^+\nu_\ell$ decays receive all  kinds of contributions.
These results  can be used to test the SU(3) flavor symmetry approach in  the four-body semileptonic $B$ decays in future experiments at LHCb and Belle-II.

\end{abstract}
\maketitle

\section{Introduction}
Semileptonic $B$ decays play a key role in testing the Standard Model  and  understanding the heavy quark dynamics.
Some three-body semileptonic $B$ decays, such as $B\to D\ell^+\nu_{\ell}$ and $B\to D^*\ell^+\nu_{\ell}$, have been  well understood. Nevertheless, other decays,  like $B\to D_{0}\ell^+\nu_{\ell}$, $B\to D^*_2\ell^+\nu_{\ell}$ and $B\to DP\ell^+\nu_\ell$ decays, received less attention.
Some branching ratios of the $B\to DP\ell'^+\nu_{\ell'}~(\ell'=e,\mu)$ decays have been measured, and the experimental data from the Particle Data Group (PDG) within $2\sigma$ errors are \cite{ParticleDataGroup:2022pth}
\begin{eqnarray}
\mathcal{B}(B^+\to D^{-}\pi^+\ell'^+\nu_{\ell'})_{T}&=& (4.4\pm0.8)\times10^{-3},\label{EBrBu2DpilvA}\\
\mathcal{B}(B^+\to D^{-}\pi^+\ell'^+\nu_{\ell'})_{\overline{D}_0}&=& (2.5\pm1.0)\times10^{-3},\label{EBrBu2Dpilv0}\\
\mathcal{B}(B^+\to D^{-}\pi^+\ell'^+\nu_{\ell'})_{\overline{D}^*_2}&=& (1.53\pm0.32)\times10^{-3},\label{EBrBu2Dpilv2}\\
\mathcal{B}(B^0\to D^{0}\pi^-\ell'^+\nu_{\ell'})_{T}&=& (4.1\pm1.0)\times10^{-3},\label{EBrBd2DpilvA}\\
\mathcal{B}(B^0\to D^{0}\pi^-\ell'^+\nu_{\ell'})_{D_0}&=& (3.0\pm2.4)\times10^{-3},\label{EBrBd2Dpilv0}\\
\mathcal{B}(B^0\to D^{0}\pi^-\ell'^+\nu_{\ell'})_{D^*_2}&=& (1.21\pm0.66)\times10^{-3},\label{EBrBd2Dpilv2}\\
\mathcal{B}(B^+\to D^{-}_sK^+\ell'^+\nu_{\ell'})_{T}&=& (3.0^{+2.8}_{-2.4})\times10^{-4},\label{EBrBu2DsKlvA}
\end{eqnarray}
where $\mathcal{B}_{T,M}$ denote the total and  $M$ resonant branching ratios. Note that, for $\mathcal{B}(B^+\to D^{-}\pi^+\ell'^+\nu_{\ell'})_{T}$ and $\mathcal{B}(B^0\to D^{0}\pi^-\ell'^+\nu_{\ell'})_{T}$ given in Eq. (\ref{EBrBu2DpilvA}) and Eq. (\ref{EBrBd2DpilvA}),  the PDG reports results only include the $D_0$ and $D^*_2$ resonances, but do not include $D^*$  resonances.
Present measurements of the $B\to DP\ell'^+\nu_{\ell'}$ decays give us an opportunity to  test theoretical approaches of the $B\to DP\ell^+\nu_{\ell}$ decays and to predict many non-measured decays, which  can be further tested  in the near future at LHCb and Belle-II.

Theoretically, semileptonic  decays are relatively simple, since the weak and strong dynamics
can be separated in these decays. All the strong dynamics in the initial and
final hadrons is included in the hadronic transition form
factors. The calculations  of the $B\to DP$ form factors are more complex than ones of the $B\to D$  form factors or the $B\to P$ form factors \cite{Feldmann:2018kqr}. So the  evaluations of the $B\to DP$ form factors  are difficult.
In the absence of reliable calculations, the symmetry analysis can provide very useful information about the decays.
SU(3) flavor symmetry is one of the popular symmetry approaches. And it  has been widely used to study  $b$-hadron decays  \cite{He:1998rq,He:2000ys,Fu:2003fy,Hsiao:2015iiu,He:2015fwa,He:2015fsa,Deshpande:1994ii,Gronau:1994rj,Gronau:1995hm,Shivashankara:2015cta,Zhou:2016jkv,Cheng:2014rfa,Wang:2021uzi,Wang:2020wxn},   $c$-hadron decays  \cite{Wang:2021uzi,Wang:2020wxn,Grossman:2012ry,Pirtskhalava:2011va,Savage:1989qr,Savage:1991wu,Altarelli:1975ye,Lu:2016ogy,Geng:2017esc,Geng:2018plk,Geng:2017mxn,Geng:2019bfz,Wang:2017azm,Wang:2019dls,Wang:2017gxe,Qiao:2024srw,Muller:2015lua}, and light hadron decays \cite{Wang:2019alu,Wang:2021uzi,Xu:2020jfr,Chang:2014iba,Zenczykowski:2005cs,Zenczykowski:2006se,Cabibbo:1963yz}.
SU(3) flavor symmetry breaking effects  due to the mass differences between the $u,d$ and $s$ quarks have also been studied, for instance,  in Refs. \cite{Martinelli:2022xir,Wang:2022fbk,Imbeault:2011jz,Wu:2005hi,Dery:2020lbc,Sasaki:2008ha,Pham:2012db,Geng:2018bow,Flores-Mendieta:1998tfv,Cheng:2012xb,Xu:2013dta,He:2014xha,Yang:2015era}.

Some four-body semileptonic decays $B/D\to P P \ell^+\nu_\ell$ and $B\to D^{(*)} P \ell^+\nu_\ell$ have been investigated  \cite{Cheng:1993ah,Tsai:2021ota,Feldmann:2018kqr,Kim:1999gm,Kim:1998nn,LeYaouanc:2018zyo,Boer:2016iez,Hambrock:2015aor,Faller:2013dwa,Shi:2021bvy,Achasov:2020qfx,Wiss:2007mr,Wang:2016wpc,Achasov:2021dvt,Gustafson:2023lrz}.
In this work,  we will  explore the $B\to DP\ell^+\nu_\ell$ decays  with  the non-resonant, the charmed  vector resonant, the charmed  scalar resonant  and  the charmed  tensor resonant contributions by the SU(3) flavor symmetry based on the experimental data.
Firstly, the hadronic amplitude relations or the form factor  relations between different decay modes will be constructed. Then, the hadronic amplitudes or the form factors will be extracted by using the available data. Finally, the not-yet-measured modes will be predicted for further tests in experiments.

This paper is organized as follows. The non-resonant contributions of the $B\to DP\ell\nu_\ell$ decays are discussed in Sec. II.
The charmed vector resonant, the charmed scalar resonant and  the charmed  tensor resonant contributions of  the $B \rightarrow  D P\ell^+\nu_\ell$ decays are presented in Sec. III.
Finally, summary is given in Sec. IV.

\section{Non-resonant $B\to D P\ell^+\nu_\ell$ decays}

\subsection{Theoretical framework}

The  non-resonant  $B \rightarrow DP \ell^+\nu_\ell$ decays  are generated by   $\bar{b}\rightarrow \bar{c} \ell^+\nu_\ell$ transition, and the effective Hamiltonian   is
\begin{eqnarray}
\mathcal{H}_{eff}(\bar{b}\rightarrow \bar{c} \ell^+\nu_\ell)=\frac{G_F}{\sqrt{2}}V_{cb}\bar{c}\gamma^\mu(1-\gamma_5)b~\bar{\nu_\ell}\gamma_\mu(1-\gamma_5)\ell,\label{Heff}
\end{eqnarray}
where $G_F$ is the Fermi constant, and $V_{cb}$ is the CKM matrix element. Decay amplitudes of the non-resonant $B \rightarrow DP \ell^+\nu_\ell$ decays can be written as
\begin{eqnarray}
\begin{aligned}
\mathcal{A}(B \rightarrow DP  \ell^+\nu_\ell)_N&=\langle D(k_1)P(k_2) \ell^+(q_1)\nu_\ell(q_2)| \mathcal{H}_{eff}(\bar{b}\rightarrow \bar{c} \ell^+\nu_\ell) |B(p_B)\rangle\\
&=\frac{G_F}{\sqrt{2}}V_{cb} L_{\mu}H^{\mu},\label{Am}
\end{aligned}
\end{eqnarray}
where  $L_\mu=\bar{\nu_\ell}\gamma_{\mu}(1-\gamma_5)\ell$ is the leptonic  charged current, and $H^\mu=\langle D(k_1)P(k_2)|\bar{c}\gamma^\mu(1-\gamma_5)b|B(p_B)\rangle$ is the hadronic amplitude.
Usually, $H^\mu$ can be obtained in terms of the form factors $F_{\bot,t,0,\|}$  of the $B \rightarrow DP$ transitions, which are are non-perturbative objects and are similar to ones of $B\to PP$ transitions \cite{Boer:2016iez}.
Nevertheless, the calculations of the $F_0,F_t,F_\perp,F_\parallel$  are very difficult. In this work, the hadronic amplitude   will be related by the SU(3) flavor symmetry.

Relevant meson multiplets are given first. Bottom pseudoscalar  triplet $B_i$,  charm pseudoscalar  triplet $D_i$, charm scalar  triplet $D_{0i}$, charm vector  triplet $D^*_i$ and  charm  tensor triplet $D^*_{2i}$  under the SU(3) flavor symmetry of the $u$, $d$, $s$ quarks are
\begin{eqnarray}
B_i &=& \Big(B^+(\bar{b} u),B^0(\bar{b} d),B^0_s(\bar{b} s)\Big),~~~~~~~~~~~~~~D_i=\Big( \overline{D}^{0}(\bar{c}u),~D^{-}(\bar{c}d),~D^{-}_s(\bar{c}s)\Big),\\
D_{0i}&=&\Big( \overline{D}^{0}_0(\bar{c}u),D^{-}_0(\bar{c}d),D^{-}_{s0}(\bar{c}s)\Big),~~~~~~~~~~~~D^*_i=\Big( \overline{D}^{*0}(\bar{c}u),~D^{*-}(\bar{c}d),~D^{*-}_s(\bar{c}s)\Big),\\
D^*_{2i}&=&\Big( \overline{D}^{*0}_2(\bar{c}u),D^{*-}_2(\bar{c}d),D^{*-}_{s2}(\bar{c}s)\Big),
\end{eqnarray}
where $i=1,2,3$ for $u$, $d$, $s$ quarks.  Note that the structures of scalar $D_0$ mesons are not known well, and they might be four-quark states,
$D_{0i}^{4q}=\Big( \overline{D}^{0}_0(\bar{c}u\bar{d}d),D^{-}_{0}(\bar{c}d\bar{u}u),D^{-}_{s0}(\bar{c}s(\bar{u}u+\bar{d}d)/\sqrt{2})\Big)$ \cite{Cheng:2003kg}.
Note that $D_{0i}$, $D^*_i$ and $D^*_{2i}$ will be used for the resonances in  Sec. \ref{RPart}, and they are given here together.
Light pseudoscalar  octets and singlets $P^i_j$ are
\begin{eqnarray}
 P^i_j&=&\left(\begin{array}{cccc}
\frac{\pi^0}{\sqrt{2}}+\frac{\eta_8}{\sqrt{6}}+\frac{\eta_1}{\sqrt{3}}& \pi^+ & K^+\\
\pi^- &-\frac{\pi^0}{\sqrt{2}}+\frac{\eta_8}{\sqrt{6}}+\frac{\eta_1}{\sqrt{3}} & K^0 \\
K^- & \overline{K}^0 &-\frac{2\eta_8}{\sqrt{6}}+\frac{\eta_1}{\sqrt{3}}\\ \end{array}\right)\,,
\end{eqnarray}
with $j=1,2,3$ for $u$, $d$, $s$ quarks. The $\eta$ and $\eta'$  are mixtures of $\eta_1=\frac{u\bar{u}+d\bar{d}+s\bar{s}}{\sqrt{3}}$ and $\eta_8=\frac{u\bar{u}+d\bar{d}-2s\bar{s}}{\sqrt{6}}$ with the mixing angle $\theta_P$
\begin{eqnarray}
\left(\begin{array}{c}
\eta\\
\eta'
\end{array}\right)\,
=
\left(\begin{array}{cc}
\mbox{cos}\theta_P&-\mbox{sin}\theta_P\\
\mbox{sin}\theta_P&\mbox{cos}\theta_P
\end{array}\right)\,\left(\begin{array}{c}
\eta_8\\
\eta_1
\end{array}\right)\,.
\end{eqnarray}\label{Eq:etamix}
And $\theta_P=[-20^\circ,-10^\circ]$ from  the PDG \cite{ParticleDataGroup:2022pth} will be used in our numerical  analysis.

The leptonic charged current is  invariant under the SU(3) flavor symmetry, and the hadronic amplitude of the non-resonant $B\to D P\ell^+\nu_\ell$ decay can be parameterized by the SU(3) flavor symmetry as
\begin{eqnarray}
H(B\to DP)_N =c_{01}B^iP_i^j{D}_j+c_{02}B^iD_iP^k_k,\label{Eq:HND2PPlv}
\end{eqnarray}
where $c_{01,02}$ are the non-perturbative coefficients under the SU(3) flavor symmetry.  The $c_{02}$ term is suppressed by the Okubo-Zweig-Iizuka (OZI) rule \cite{Okubo:1963fa,Lipkin:1986bi,Lipkin:1996ny}, and it only appears in the decays with $\eta,\eta'$ final states.
The idiographic hadronic amplitudes of the  non-resonant $B\to DP\ell^+\nu_\ell$ decays  are
given in  Tab. \ref{Tab:HD2DPlvAmp}.   From Tab. \ref{Tab:HD2DPlvAmp}, one can see that,  if ignoring the OZI suppressed $c_{02}$ contributions,  all hadronic amplitudes  may be related by the nonperturbative coefficient $c_{01}$.

\begin{table}[t]
\renewcommand\arraystretch{1.5}
\tabcolsep 0.2in
\centering
\caption{The hadronic amplitudes for the non-resonant $B\to DP\ell^+\nu$ decays under the SU(3) flavor symmetry. }\vspace{0.1cm}
{\footnotesize
\begin{tabular}{lc|lc}  \hline\hline
~~~Decay modes~~~~~  & Hadronic amplitudes  &~~~Decay modes~~~~~  & Hadronic amplitudes   \\\hline
%
$B^+\to D^-\pi^+\ell^+\nu_\ell$                       &$c_{01}$                                                                                    &$B^0\to D^-\pi^0\ell^+\nu_\ell$                         &$-\frac{1}{\sqrt{2}}c_{01}$                                                                 \\\hline
$B^+\to D^-_sK^+\ell^+\nu_\ell$                       &$c_{01}$                                                                                    &$B^0\to D^-\eta\ell^+\nu_\ell$                          &$\frac{c_{01}cos\theta_P}{\sqrt{6}}-\frac{(c_{01}+3c_{02})sin\theta_P}{\sqrt{3}}$           \\\hline
$B^+\to \overline{D}^0\pi^0\ell^+\nu_\ell$            &$\frac{1}{\sqrt{2}}c_{01}$                                                                  &$B^0\to D^-\eta'\ell^+\nu_\ell$                         &$\frac{c_{01}sin\theta_P}{\sqrt{6}}+\frac{(c_{01}+3c_{02})cos\theta_P}{\sqrt{3}}$           \\\hline
$B^+\to \overline{D}^0\eta\ell^+\nu_\ell$             &$\frac{c_{01}cos\theta_P}{\sqrt{6}}-\frac{(c_{01}+3c_{02})sin\theta_P}{\sqrt{3}}$           &$B^0_s\to \overline{D}^0K^-\ell^+\nu_\ell$              &$c_{01}$                                                                                    \\\hline
$B^+\to \overline{D}^0\eta'\ell^+\nu_\ell$            &$\frac{c_{01}sin\theta_P}{\sqrt{6}}+\frac{(c_{01}+3c_{02})cos\theta_P}{\sqrt{3}}$           &$B^0_s\to D^-\overline{K}^0\ell^+\nu_\ell$              &$c_{01}$                                                                                    \\\hline
$B^0\to \overline{D}^0\pi^-\ell^+\nu_\ell$              &$c_{01}$                                                                                    &$B^0_s\to D_s^-\eta\ell^+\nu_\ell$                      &$-\frac{2c_{01}cos\theta_P}{\sqrt{6}}-\frac{(c_{01}+3c_{02})sin\theta_P}{\sqrt{3}}$         \\\hline
$B^0\to D^-_sK^0\ell^+\nu_\ell$                         &$c_{01}$                                                                                    &$B^0_s\to D_s^-\eta'\ell^+\nu_\ell$                     &$-\frac{2c_{01}sin\theta_P}{\sqrt{6}}+\frac{(c_{01}+3c_{02})cos\theta_P}{\sqrt{3}}$         \\\hline
\end{tabular}\label{Tab:HD2DPlvAmp}}                                                                                                                 %
\end{table}

The differential branching ratios of the non-resonant $B \rightarrow DP \ell^+\nu_\ell$ decays are \cite{Boer:2016iez}
\begin{eqnarray}
    \frac{d\mathcal{B}(B \rightarrow DP  \ell^+\nu_\ell)_N}{dq^2 dk^2}=\frac{1}{2}\tau_{B}|\mathcal{N}|^2\beta_\ell(3-\beta_\ell)|H_N|^2,
\end{eqnarray}
with
\begin{eqnarray}
{|\mathcal{N}|}^2&=&G^2_F {|V_{cb}|}^2 \frac{\beta_\ell q^2 \sqrt{\lambda}}{3 \times 2^{10} \pi^5 m^3_B}\,,\quad \text{with}\quad
\beta_\ell=1-\frac{m^2_\ell}{q^2}\,. \nonumber \\
|H_N|^2&=& |F_0|^2+\frac{2}{3}(|F_{\parallel}|^2+|F_\perp|^2)+\frac{3m_\ell^2}{q^2(3-\beta_\ell)}|F_t|^2,\label{Eq:DB2PPlvdbr}
\end{eqnarray}
where $\tau_{M}$($m_{M}$) is lifetime(mass) of $M$ particle. The ranges of integration are given by  $(m_{D}+m_{P})^2\leq k^2\leq(m_{B}-m_\ell)^2$ and $m_\ell^2\leq q^2 \leq(m_{B}-\sqrt{k^2})^2$.
 If we ignore $|F_t|^2$ term since it is proportional to $m_\ell^2$ and it is small when $\ell=e,\mu$, $|H_N|^2$ is only include the hadronic part.
Noted that although $|F_t|^2$ term might be large when $\ell=\tau$, it is difficult to estimate its contribution in this work, so we still ignore it.
Then $H_N$, which only includes hadronic part, follow the relationship of the SU(3) flavor symmetry in Tab. \ref{Tab:HD2DPlvAmp}.

\subsection{Numerical results}
For the non-resonant $B\to DP\ell^+\nu_{\ell}$ decays, no any  process has been measured until now. However, as given in Eq. (\ref{EBrBu2DsKlvA}), $\mathcal{B}(B^+\to D^{-}_sK^+\ell'^+\nu_{\ell'})_{T}$ has been measured. The  $B^+\to D^{-}_sK^+\ell'^+\nu_{\ell'}$ mode can decay via the non-resonance and the $D^*_2$  tensor meson resonance. In the subsequent analysis in  Sec. \ref{RPart},  the contributions of $D^*_2$  tensor meson resonance are far less than the experimental data given in Eq. (\ref{EBrBu2DsKlvA}). So we think that the  non-resonant contributions are dominant  in the $B^+\to D^{-}_sK^+\ell'^+\nu_{\ell'}$ decays, $i.e.$, $\mathcal{B}(B^+\to D^{-}_sK^+\ell'^+\nu_{\ell'})_{N}\approx\mathcal{B}(B^+\to D^{-}_sK^+\ell'^+\nu_{\ell'})_{T}$.
 The experimental data of $\mathcal{B}(B^+\to D^{-}_sK^+\ell'^+\nu_{\ell'})_{T}$ are used to determine $c_{01}$ in the non-resonant $B\to DP\ell^+\nu_{\ell'}$ decays (Due to poor relevant  experimental data, the OZI suppressed $c_{02}$ term is ignored). Then many other branching ratios of the non-resonant $B\to DP\ell^+\nu_{\ell}$  decays can be predicted by using the data of $\mathcal{B}(B^+\to D^{-}_sK^+\ell'^+\nu_{\ell'})_{T}$, which are listed in the second column of Tab.  \ref{Tab:BrD2DPlv}.

From the second column of Tab.  \ref{Tab:BrD2DPlv}, one can see that many branching ratio central values of the non-resonant $B\to DP\ell'^+\nu_{\ell'}$ decays, such as
$\mathcal{B}(B^+\to D^-\pi^+\ell'^+\nu_{\ell'})_N$, $\mathcal{B}(B^+\to \overline{D}^0\pi^0\ell'^+\nu_{\ell'})_N$, $\mathcal{B}(B^+\to \overline{D}^0\eta\ell'^+\nu_{\ell'})_N$, $\mathcal{B}(B^0\to \overline{D}^0\pi^-\ell'^+\nu_{\ell'}$
$\mathcal{B}(B^0\to D^-_sK^0\ell'^+\nu_{\ell'})_N$, $\mathcal{B}(B^0\to D^-\pi^0\ell'^+\nu_{\ell'})_N$, $\mathcal{B}(B^0\to D^-\eta\ell'^+\nu_{\ell'})_N$, $\mathcal{B}(B^0_s\to \overline{D}^0K^-\ell'^+\nu_{\ell'})_N$
$\mathcal{B}(B^0_s\to D^-\overline{K}^0\ell'^+\nu_{\ell'})_N$ and  $\mathcal{B}(B^0_s\to D_s^-\eta\ell'^+\nu_{\ell'})_N$,  are on the orders
of $10^{-4}$, which could be measured by the LHCb and Belle II experiments. Nevertheless, other decays, for example, the non-resonant
$B^+\to \overline{D}^0\eta'\ell'^+\nu_{\ell'}$, $B^0\to D^-\eta'\ell'^+\nu_{\ell'}$, $B^0_s\to D_s^-\eta'\ell'^+\nu_{\ell'}$, and  all $B\to DP\tau^+\nu_{\tau}$  decays,
are strongly suppressed by the narrow phase spaces, their
branching ratio central values are on the orders of  $\mathcal{O}(10^{-5}-10^{-7})$, and they might not be observed by the experiments in the near future.

\begin{table}[hb]
\renewcommand\arraystretch{1.35}
\tabcolsep 0.15in
\centering
\caption{Branching ratios  for the $B\to DP\ell^+\nu_\ell$ decays within  $2\sigma$ errors (in units of  $10^{-3}$).
$\mathcal{B}_{N}$ denotes the non-resonant branching ratios,  $\mathcal{B}_{[R]}$ denotes the $R$ resonant ones,  $^e$denotes experimental data within $2\sigma$ errors,
and  $^\dag$denotes the results obtained by considering the resonance width effects.
}\vspace{0.05cm}
{\footnotesize
\begin{tabular}{lcccc}  \hline\hline
~~~~~Decay modes~~~~~                                       &  $\mathcal{B}_{N}$                               &~~~~~~~~~$\mathcal{B}_{[D^*]}$                             &~~~~~~~~$\mathcal{B}_{[D_0]}$                                                          &~~~~~~~~~~~~~~ $\mathcal{B}_{[D^*_2]}$                                     \\\hline
$B^+\to D^-\pi^+\ell'^+\nu_{\ell'}$                       &$0.64\pm0.52$                                      &$\cdots$                                                    &$^{2.64\pm0.86_{[D^0_0]}}_{2.5\pm1.0^{e}}$                                         &$^{1.33\pm0.12_{[D^{*0}_2]}}_{1.53\pm0.32^{e}}$                         \\\hline
$B^+\to D^-_sK^+\ell'^+\nu_{\ell'}$                       &$^{0.32\pm0.26}_{0.30^{+0.28e}_{-0.24}}$          &$\cdots$                                                     &$\cdots$                                                                            &$[5.66\times10^{-15},2.15\times10^{-7}]$$_{[D^{*0}_2]}$                 \\\hline
$B^+\to \overline{D}^0\pi^0\ell'^+\nu_{\ell'}$            &$0.32\pm0.26$                                      &$^{35.10\pm2.68_{[D^{*0}]}}_{31.22\pm2.25^\dag_{[D^{*0}]}}$      &$1.34\pm0.44_{[D^0_0]}$                                                            &$0.70\pm0.06$$_{[D^{*0}_2]}$                                            \\\hline
$B^+\to \overline{D}^0\eta\ell'^+\nu_{\ell'}$             &$0.11\pm0.10$                                      &$\cdots$                                                    &$\cdots$                                                                           &$(4.36\pm1.22)\times10^{-3}$$_{[D^{*0}_2]}$                             \\\hline
$B^+\to \overline{D}^0\eta'\ell'^+\nu_{\ell'}$            &$0.038\pm0.033$                                    &$\cdots$                                                    &$\cdots$                                                                           &$\cdots$                                                                \\\hline
$B^0\to \overline{D}^0\pi^-\ell'^+\nu_{\ell'}$              &$0.60\pm0.49$                                    &$^{33.67\pm2.11_{[D^{*-}]}}_{30.14\pm1.72^\dag_{[D^{*-}]}}$      &$^{2.48\pm0.81_{[D^-_0]}}_{3.0\pm2.4^e}$                                           &$^{1.28\pm0.13_{[D^{*-}_2]}}_{1.21\pm0.66^e}$                           \\\hline
$B^0\to D^-_sK^0\ell'^+\nu_{\ell'}$                         &$0.30\pm0.24$                                    &$\cdots$                                                    &$\cdots$                                                                           &$[4.07\times10^{-14},5.30\times10^{-6}]$$_{[D^{*-}_2]}$                \\\hline
$B^0\to D^-\pi^0\ell'^+\nu_{\ell'}$                         &$0.30\pm0.24$                                    &$^{15.29\pm1.21_{[D^{*-}]}}_{13.69\pm1.05^\dag_{[D^{*-}]}}$      &$1.23\pm0.40_{[D^-_0]}$                                                            &$0.62\pm0.06$$_{[D^{*-}_2]}$                                           \\\hline
$B^0\to D^-\eta\ell'^+\nu_{\ell'}$                          &$0.10\pm0.09$                                    &$\cdots$                                                    &$\cdots$                                                                           &$(3.54\pm1.40)\times10^{-3}$$_{[D^{*-}_2]}$                            \\\hline
$B^0\to D^-\eta'\ell'^+\nu_{\ell'}$                         &$0.035\pm0.031$                                  &$\cdots$                                                    &$\cdots$                                                                           &$\cdots$                                                               \\\hline
$B^0_s\to \overline{D}^0K^-\ell'^+\nu_{\ell'}$              &$0.41\pm0.33$                                    &$\cdots$                                                    &$\cdots$                                                                           &$1.23\pm0.18$$_{[D^{*-}_{s2}]}$                                        \\\hline
$B^0_s\to D^-\overline{K}^0\ell'^+\nu_{\ell'}$              &$0.40\pm0.33$                                    &$\cdots$                                                    &$\cdots$                                                                           &$1.11\pm0.16$$_{[D^{*-}_{s2}]}$                                        \\\hline
$B^0_s\to D_s^-\eta\ell'^+\nu_{\ell'}$                      &$0.15\pm0.13$                                    &$\cdots$                                                    &$\cdots$                                                                           &$(1.67\pm0.59)\times10^{-2}$$_{[D^{*-}_{s2}]}$                           \\\hline
$B^0_s\to D_s^-\eta'\ell'^+\nu_{\ell'}$                     &$0.095\pm0.081$                                  &$\cdots$                                                    &$\cdots$                                                                           &$\cdots$                                                                 \\\hline
$B^+\to D^-\pi^+\tau^+\nu_\tau$                          &$0.091\pm0.074$                                    &$\cdots$                                                     &$0.35\pm0.12_{[D^0_0]}$                                                            &$(8.62\pm1.82)\times10^{-2}$$_{[D^{*0}_2]}$                                             \\\hline
$B^+\to D^-_sK^+\tau^+\nu_\tau$                          &$0.022\pm0.017$                                    &$\cdots$                                                     &$\cdots$                                                                           &$[3.54\times10^{-16},1.45\times10^{-8}]$$_{[D^{*0}_2]}$                  \\\hline
$B^+\to \overline{D}^0\pi^0\tau^+\nu_\tau$               &$0.047\pm0.038$                                    &$^{8.59\pm0.66_{[D^{*0}]}}_{7.63\pm0.55^\dag_{[D^{*0}]}}$                                     &$0.18\pm0.06_{[D^0_0]}$                                                            &$(4.52\pm0.95)\times10^{-2}$$_{[D^{*0}_2]}$                                             \\\hline
$B^+\to \overline{D}^0\eta\tau^+\nu_\tau$                &$0.0085\pm0.0072$                                  &$\cdots$                                                     &$\cdots$                                                                           &$(2.81\pm1.00)\times10^{-4}$$_{[D^{*0}_2]}$                              \\\hline
$B^+\to \overline{D}^0\eta'\tau^+\nu_\tau$               &$0.00086\pm0.00074$                                &$\cdots$                                                     &$\cdots$                                                                           &$\cdots$                                                                \\\hline
$B^0\to \overline{D}^0\pi^-\tau^+\nu_\tau$                 &$0.086\pm0.070$                                    &$^{8.22\pm0.52_{[D^{*-}]}}_{7.34\pm0.42^\dag_{[D^{*-}]}}$                                   &$0.32\pm0.11_{[D^-_0]}$                                                            &$(8.34\pm1.78)\times10^{-2}$$_{[D^{*-}_2]}$                                            \\\hline
$B^0\to D^-_sK^0\tau^+\nu_\tau$                            &$0.020\pm0.016$                                    &$\cdots$                                                   &$\cdots$                                                                           &$[2.51\times10^{-15},3.20\times10^{-7}]$$_{[D^{*-}_2]}$                 \\\hline
$B^0\to D^-\pi^0\tau^+\nu_\tau$                            &$0.043\pm0.035$                                    &$^{3.73\pm0.30_{[D^{*-}]}}_{3.34\pm0.26^\dag_{[D^{*-}]}}$                                   &$0.16\pm0.06_{[D^-_0]}$                                                            &$(4.06\pm0.87)\times10^{-2}$$_{[D^{*-}_2]}$                                            \\\hline
$B^0\to D^-\eta\tau^+\nu_\tau$                             &$0.0077\pm0.0066$                                  &$\cdots$                                                   &$\cdots$                                                                           &$(2.34\pm1.03)\times10^{-4}$$_{[D^{*-}_2]}$                             \\\hline
$B^0\to D^-\eta'\tau^+\nu_\tau$                            &$0.00077\pm0.00067$                                &$\cdots$                                                   &$\cdots$                                                                           &$\cdots$                                                                \\\hline
$B^0_s\to \overline{D}^0K^-\tau^+\nu_\tau$                 &$0.040\pm0.033$                                    &$\cdots$                                                   &$\cdots$                                                                           &$(7.86\pm1.99)\times10^{-2}$$_{[D^{*-}_{s2}]}$                                        \\\hline
$B^0_s\to D^-\overline{K}^0\tau^+\nu_\tau$                 &$0.039\pm0.032$                                    &$\cdots$                                                   &$\cdots$                                                                           &$(7.10\pm1.79)\times10^{-2}$$_{[D^{*-}_{s2}]}$                                        \\\hline
%
%
$B^0_s\to D_s^-\eta\tau^+\nu_\tau$                         &$0.011\pm0.009$                                    &$\cdots$                                                   &$\cdots$                                                                           &$(1.04\pm0.44)\times10^{-3}$$_{[D^{*-}_{s2}]}$                          \\\hline
$B^0_s\to D_s^-\eta'\tau^+\nu_\tau$                        &$0.0020\pm0.0017$                                  &$\cdots$                                                   &$\cdots$                                                                           &$\cdots$                                                                  \\\hline
\end{tabular}\label{Tab:BrD2DPlv}}
\end{table}

\clearpage

\section{Decays $B\to D P\ell^+\nu_\ell$  with the $D^*,D_0,D^*_2$  resonances  }\label{RPart}

For the  $B\to D P\ell^+\nu_\ell$ decays with the resonances,
when the decay widths of the resonance states are very narrow,  the resonant branching ratios respect a simple factorization relation
\begin{equation}
\mathcal{B}(B \rightarrow D_J \ell^+\nu_\ell, D_J \rightarrow D P) = \mathcal{B}(B \rightarrow D_J \ell^+\nu_\ell) \times  \mathcal{B}(D_J \rightarrow D P),\label{Eq:Br4BD}
\end{equation}
due to parity conservation, only $D_J=D_0/D^*/D^*_2$ resonances are considered for the $B\to DP\ell^+\nu_\ell$ decays \cite{Kim:1998np}.
And this result is also a good approximation for wider resonances. Eq. (\ref{Eq:Br4BD}) will be used in our analysis for resonant $B \rightarrow D_J(\rightarrow D P) \ell^+\nu_\ell$ decays.
We first calculate  $\mathcal{B}(B \rightarrow D_J \ell^+\nu_\ell)$ and $\mathcal{B}(D_J \rightarrow D P)$ by the SU(3) flavor symmetry, then obtain $\mathcal{B}(B \rightarrow D_J \ell^+\nu_\ell, D_J \rightarrow D P)$.
The width effects of some  resonance  states will be analyzed later.

\subsection{Decays $B\to D_J\ell^+\nu_\ell$ }
The effective  Hamiltonian  is given in Eq. (\ref{Heff}),  and the amplitudes of the $B\to D_J\ell^+\nu_\ell$ decays also can
be factorized in the product of the matrix elements of
leptonic and hadronic currents.
The differential branching ratios  of the  $B\to D_J \ell^+\nu_\ell$ decays are  \cite{Ivanov:2019nqd}
\begin{equation}
 \frac{d\mathcal{B}(B \to
  {D_J}\ell^+\nu_\ell)}{dq^2}=\frac{\tau_{B}G_F^2 |V_{cb}|^2\lambda^{1/2}(q^2-m_\ell^2)^2}{24(2\pi)^3m_{B}^3q^2}
 {\cal H}_{\rm total},\label{Eq:dbdq2}
\end{equation}
with
\begin{eqnarray}
  \label{eq:htot}
 {\cal H}_{\rm total}&=& ({\cal H}_U+{\cal
   H}_L)\left(1+\frac{m_\ell^2}{2q^2}\right) +\frac{3m_\ell^2}{2q^2}{\cal H}_S,\\
   {\cal H}_U&=&|H^{D_J}_+|^2+|H^{D_J}_-|^2, ~~~ {\cal H}_L=|H^{D_J}_0|^2, ~~~
    {\cal H}_S=|H^{D_J}_t|^2,  \label{eq:hh}
\end{eqnarray}
where  $\lambda\equiv
\lambda(m_{B}^2,m_{D_J}^2,q^2)$  with $\lambda(a,b,c)=a^2+b^2+c^2-2ab-2ac-2bc$, and $m_\ell^2\leq q^2\leq(m_{B}-m_{D_J})^2$.  The hadronic helicity amplitudes are defined through the hadronic matrix elements
\begin{eqnarray}
H^{D_J}_{m'}&=&\epsilon^*_{\beta}(m')\langle{D_J}(p,\varepsilon^*)|\bar{c}\gamma^\beta(1-\gamma_5)b|B(p_B)\rangle,\label{Eq:HDJm}
\end{eqnarray}
where  $\varepsilon^*$ is the polarization of the $D^*$ and $D^*_2$ mesons, and $\epsilon_{\mu}(m)$  is the polarization  of the virtual $W$ with $m=0,t,\pm1$.

The hadronic helicity amplitudes related to the form factors  are
\begin{eqnarray}
H^D_\pm=0,~~~~~~~~~~~~~~~~~~H^D_0=\frac{2m_{B}|\vec{p}_{D}|}{\sqrt{q^2}}f^{BD}_+(q^2),~~~~~~~~~~~~~~~~~~H^D_t=\frac{m_{B}^2-m_{D}^2}{\sqrt{q^2}}f^{BD}_0(q^2),
\end{eqnarray}
for $B\to D\ell^+\nu_\ell$ decays,
\begin{eqnarray}
H^{D_0}_\pm&=&0,~~~~~~~~~~~~~~~H^{D_0}_0=\frac{i2m_{B}|\vec{p}_{D_0}|}{\sqrt{q^2}}f^{BD_0}_1(q^2),~~~~~~~~~~~~~~~H^{D_0}_t=\frac{i (m_{B}^2-m_{D_0}^2)}{\sqrt{q^2}}f^{BD_0}_0(q^2),
\end{eqnarray}
for $B\to D_0\ell^+\nu_\ell$ decays,
\begin{eqnarray}
H^{D^*}_{\pm}&=&(m_{B}+m_{D^*})A^{BD^*}_1(q^2)\mp\frac{2m_{B}|\vec{p}_{D^*}|}{(m_{B}+m_{D^*})}V^{BD^*}(q^2), \\
H^{D^*}_{0}&=& \frac{1}{2m_{D^*}\sqrt{q^2}}\left[(m_{B}^2-m_{D^*}^2-q^2)(m_{B}+m_{D^*})A^{BD^*}_1(q^2)-\frac{4m_{B}^2|\vec{p}_{D^*}|^2}{m_{B}+m_{D^*}}A^{BD^*}_2(q^2)\right], \\
H^{D^*}_{t}&=& \frac{2m_{B}|\vec{p}_{D^*}|}{\sqrt{q^2}}A^{BD^*}_0(q^2),
\end{eqnarray}
for $B\to D^*\ell^+\nu_\ell$ decays, and
\begin{eqnarray}
H^{D^*_2}_{\pm} &=&\frac{2|\vec{p}_{D^*_2}|}{\sqrt{6}m_{D^*_2}}\left[(m_{B}+m_{D^*_2})A^{BD^*_2}_1(q^2)\mp\frac{2m_{B}|\vec{p}_{D^*_2}|}{(m_{B}+m_{D^*_2})}V^{BD^*_2}(q^2)\right], \\
H^{D^*_2}_{0}&=& \frac{|\vec{p}_{D^*_2}|}{\sqrt{2}m_{D^*_2}}\frac{1}{2m_{D^*_2}\sqrt{q^2}}\left[(m_{B}^2-m_{D^*_2}^2-q^2)(m_{B}+m_{D^*_2})A^{BD^*_2}_1(q^2)-\frac{4m_{B}^2|\vec{p}_{D^*_2}|^2}{m_{B}+m_{D^*_2}}A^{BD^*_2}_2(q^2)\right], \\
H^{D^*_2}_{t}&=& \frac{|\vec{p}_{D^*_2}|}{\sqrt{2}m_{D^*_2}}\frac{2m_{B}|\vec{p}_{D^*_2}|}{\sqrt{q^2}}A^{BD^*_2}_0(q^2),
\end{eqnarray}
for $B\to D^*_2\ell^+\nu_\ell$ decays,  where $|\vec{p}_{D_J}|\equiv\sqrt{\lambda(m_{B}^2,m_{D_J}^2,q^2)}/(2m_{B})$.

The form factors of the  $B\to D_J$ transitions   are defined as   \cite{Cheng:2017pcq,Cheng:2003sm}
\begin{eqnarray}
\left<D(p)\left|\bar{c}\gamma_{\mu}b\right|B(p_B)\right>&=&f_1^{BD}(q^2)\Big((p+p_B)_{\mu}-\frac{m_{B}^2-m_{D}^2}{q^2}q_\mu\Big)+f_0^{BD}(q^2)\frac{m_{B}^2-m_{D}^2}{q^2}q_\mu ,\\
\left<D_0(p)\left|\bar{c}\gamma_{\mu}\gamma_5b\right|B(p_B)\right>&=&-i\Bigg[f_1^{BD_0}(q^2)\Big((p+p_B)_{\mu}-\frac{m_{B}^2-m_{D_0}^2}{q^2}q_\mu\Big)+f_0^{BD_0}(q^2)\frac{m_{B}^2-m_{D_0}^2}{q^2}q_\mu \Bigg],\\
\left<D^*(p,\varepsilon^*)\left|\bar{c}\gamma_{\mu}(1-\gamma_5)b\right|B(p_B)\right>
&=&\frac{2V^{BD^*}(q^2)}{m_B+m_{D^*}}\epsilon_{\mu\nu\alpha\beta}\varepsilon^{*\nu}p^\alpha_Bp^\beta\nonumber\\
&&-i\left[\varepsilon^*_\mu(m_B+m_{D^*})A^{BD^*}_1(q^2)-(p_B+p)_\mu(\varepsilon^*.p_B)\frac{A^{BD^*}_2(q^2)}{m_B+m_{D^*}}\right]\nonumber\\
&&+iq_\mu(\varepsilon^*.p_B)\frac{2m_{D^*}}{q^2}\Big[A^{BD^*}_3(q^2)-A^{BD^*}_0(q^2)\Big],\\
%
\left<D^*_2(p,\varepsilon^{*})\left|\bar{c}\gamma_{\mu}(1-\gamma_5)b\right|B(p_B)\right>
&=&\frac{2iV^{BD^*_2}(q^2)}{m_B+m_{D^*_2}}\epsilon_{\mu\nu\alpha\beta}e^{*\nu}p^\alpha_Bp^\beta\nonumber\\
&&+2m_{D^*_2}\frac{e^*\cdot q}{q^2}q_\mu A^{BD^*_2}_0(q^2)+(m_B+m_{D^*_2})\Big(e^*_\mu-\frac{e^*\cdot q}{q^2}q_\mu\Big)A^{BD^*_2}_1(q^2)\nonumber\\
&&-\frac{e^*\cdot q}{m_B+m_{D^*_2}}\Big((p_B+p)_\mu-\frac{m_B^2-m^2_{D^*_2}}{q^2}q_\mu\Big)A^{BD^*_2}_2(q^2),
\end{eqnarray}
where $s=q^2$ ($q=p_B-p$)    and $e^{*\nu}\equiv \frac{\varepsilon^{*\mu\nu}\cdot p_{B\mu}}{m_B}$.

Now one can obtain the  branching ratios  of the  $B\to D_J \ell^+\nu_\ell$ decays by the relevant form factors, which  depend on the  different methods.
In this work, we use the SU(3) flavor symmetry to obtain the relations of the hadronic amplitudes, and the same relations are also true for the form factors.
In terms of the SU(3) flavor symmetry, the hadronic helicity amplitudes defined in Eq. (\ref{Eq:HDJm}) can be parameterized as
\begin{eqnarray}  \label{eq:htot}
H^{D_J}_{m'}=C^{D_J}_0B_i(D_J)^i,
\end{eqnarray}
where $C^{D_J}_0$ are the non-perturbative coefficients under  the SU(3)
flavor symmetry. For the charmed  four-quark states $D_0^{4q}$, $H^{D_0}_{m'}=C^{4q,D_0}_0B_i(D_J)^{ij}_j$.  And the hadronic amplitude  relations for the $B\to D_J \ell^+\nu_\ell$ decays are summarized in Tab. \ref{Tab:HB2DJlvH}.

\begin{table}[ht]
\renewcommand\arraystretch{1.0}
\tabcolsep 0.2in
\centering
\caption{The hadronic amplitudes for  $B\to D_{J}\ell^+\nu$ decays  under the SU(3) flavor symmetry. }\vspace{0.1cm}
\begin{tabular}{cc|cc}  \hline\hline
Decay moeds                                &  SU(3) hadronic amplitudes                        &Decay moeds                                    & SU(3) hadronic amplitudes     \\\hline
$B^+\to \overline{D}^0\ell^+\nu_\ell$    &$C^D_{0}$                                          &$B^+\to \overline{D}^{*0}\ell^+\nu_\ell$     &$C^{D^*}_{0}$                   \\
$B^0\to D^-\ell^+\nu_\ell$               &$C^D_{0}$                                          &$B^0\to D^{*-}\ell^+\nu_\ell$                &$C^{D^*}_{0}$                   \\
$B^0_s\to D^-_s\ell^+\nu_\ell$             &$C^D_{0}$                                          &$B^0_s\to D^{*-}_s\ell^+\nu_\ell$              &$C^{D^*}_{0}$                   \\\hline
$B^+\to \overline{D}^0_0\ell^+\nu_\ell$  &$C^{D_0}_{0},~~~~C^{4q,D_0}_{0}$                   &$B^+\to \overline{D}^{*0}_2\ell^+\nu_\ell$   &$C^{D^*_2}_{0}$                  \\
$B^0\to D^-_0\ell^+\nu_\ell$             &$C^{D_0}_{0},~~~~C^{4q,D_0}_{0}$                   &$B^0\to D^{*-}_2\ell^+\nu_\ell$              &$C^{D^*_2}_{0}$                    \\
$B^0_s\to D^-_{s0}\ell^+\nu_\ell$          &$C^{D_0}_{0},~~\sqrt{2}C^{4q,D_0}_{0}$             &$B^0_s\to D^{*-}_{s2}\ell^+\nu_\ell$           &$C^{D^*_2}_{0}$                    \\\hline
\end{tabular}\label{Tab:HB2DJlvH}
\end{table}

The relations in Tab. \ref{Tab:HB2DJlvH} will be used for the form factors  $F(0)$, which are $f^{BD}_{1}(0)$ in the $B\to D\ell^+\nu_\ell$ decays, $V^{BD^*}(0)$  in the $B\to D^{*}\ell^+\nu_\ell$ decays,  $f^{BD_0}_{1}(0)$ in the $B\to D_0\ell^+\nu_\ell$ decays, and $A^{BD^*_2}_1(0)$ in $B\to D^*_2\ell^+\nu_\ell$ decays. The form factors  $F(0)$ are determined by the relevant experimental data.   Other form factors $F_i(0)$ can be expressed as $r_i\times F(0)$,  and the values of the ratios $r_i=\frac{F_i(0)}{F(0)}$ are taken from Ref. \cite{Totten:2000ab} for the $B\to D/D^*\ell^+\nu_\ell$ decays, from Ref. \cite{Verma:2011yw} for the $B\to D_0\ell^+\nu_\ell$ decays and from
Ref. \cite{Chen:2021ywv} for the $B\to D^*_2\ell^+\nu_\ell$ decays. Taking the $B\to D^{*}\ell^+\nu_\ell$ decays as an example, there are  four form factors $V^{BD^*}(0)$ and $A^{BD^*}_{0,1,2}(0)$ in the $B\to D^{*}\ell^+\nu_\ell$ decays,
 $A^{BD^*}_{0,1,2}(0)$ are expressed by $r_{0,1,2}\times V^{BD^*}(0)$, and the values of $r_{0,1,2}=\frac{A^{BD^*}_{0,1,2}(0)}{V^{BD^*}(0)}$ are taken from Ref. \cite{Totten:2000ab}, and then there is  only one parameter $V^{BD^*}(0)$ in the $B\to D^{*}\ell^+\nu_\ell$ decays, and it can be determined by the experimental data of the $B\to D^{*}\ell^+\nu_\ell$ decays.

Now we give our branching ratio predictions of the  semileptonic $B\to D_J\ell^+\nu_\ell$  decays under the SU(3) flavor symmetry. If not specially specified, the theoretical input
parameters, such as the lifetimes, the masses, and the
experimental data within the $2\sigma$ error bars from PDG \cite{ParticleDataGroup:2022pth}
will be used in our numerical analysis.

Theoretically, exclusive semileptonic $B \to D /D^* \ell^+\nu_\ell$ are well understood.
Although the $B\to D \ell^+\nu_\ell$ decays are not used for the four-body semileptonic decay branching ratios, there are five  experimental data in  the $B\to D\ell^+\nu_\ell$ decays, which could be used to test the SU(3) flavor symmetry approach, so we present their results here.
The experimental data of the $B\to D \ell^+\nu_\ell$ decays are listed in the second column of Tab. \ref{Tab:BrB2DDVlv}, which are used to constrain the only one free parameter $f^{BD}_1(0)$.
We obtain  that $f^{BD}_1(0)=0.66\pm0.05$, which agrees with $0.67$ given in Ref. \cite{Totten:2000ab}.  Then one can predict the  branching ratios of the $B^0_s\to D^-_s\ell^+\nu_{\ell}$  decays in terms of the constrained $f^{BD}_1(0)$, which  are listed in the third column of Tab. \ref{Tab:BrB2DDVlv}.

\begin{table}[htbp]
\renewcommand\arraystretch{1.2}
\tabcolsep 0.15in
\centering
\caption{The experimental data and the SU(3) flavor symmetry predictions of $\mathcal{B}(B\to D/D^*\ell^+\nu_\ell)$ within $2\sigma$ errors (in units of $10^{-2}$).
$^a$denotes that the experimental data are not used to constrain the parameter $C^{D^*}_{0}$.
}\vspace{0.1cm}
{\footnotesize
\begin{tabular}{lcc|lcc}  \hline\hline
Branching ratios                                         & Exp. data \cite{ParticleDataGroup:2022pth}    & Our predictions              & Branching ratios  & Exp. data \cite{ParticleDataGroup:2022pth} & Our predictions \\\hline
$\mathcal{B}(B^+\to \overline{D}^0\ell'^+\nu_{\ell'})$     &$2.30\pm0.18$               &$2.34\pm0.14$                               & $\mathcal{B}(B^+\to \overline{D}^{*0}\ell'^+\nu_{\ell'})$   &$5.58\pm0.44$               &$5.41\pm0.27$\\
$\mathcal{B}(B^0\to D^-\ell'^+\nu_{\ell'})$                &$2.24\pm0.18$               &$2.19\pm0.13$                               & $\mathcal{B}(B^0\to D^{*-}\ell'^+\nu_{\ell'})$              &$4.97\pm0.24$               &$4.97\pm0.24$\\
$\mathcal{B}(B^0_s\to D^-_s\ell'^+\nu_{\ell'})$              &$\cdots$                     &$2.20\pm0.14$                               & $\mathcal{B}(B^0_s\to D^{*-}_s\ell'^+\nu_{\ell'})$           &$\cdots$                    &$4.99\pm0.28$\\
$\mathcal{B}(B^0_s\to D^-_s\mu^+\nu_\mu)$                    &$2.44\pm0.46$               &$2.20\pm0.14$                               & $\mathcal{B}(B^0_s\to D^{*-}_s\mu^+\nu_\mu)$                  &$5.30\pm1.0$                &$4.98\pm0.28$\\
$\mathcal{B}(B^+\to \overline{D}^0\tau^+\nu_\tau)$         &$0.77\pm0.50$               &$0.68\pm0.04$                               & $\mathcal{B}(B^+\to \overline{D}^{*0}\tau^+\nu_\tau)$       &$1.88\pm0.40^a$               &$1.35\pm0.07$\\
$\mathcal{B}(B^0\to D^-\tau^+\nu_\tau)$                    &$1.05\pm0.46$               &$0.64\pm0.04$                               & $\mathcal{B}(B^0\to D^{*-}\tau^+\nu_\tau)$                  &$1.58\pm0.18^a$               &$1.21\pm0.06$\\
$\mathcal{B}(B^0_s\to D^-_s\tau^+\nu_\tau)$                  &$\cdots$                    &$0.63\pm0.04$                               & $\mathcal{B}(B^0_s\to D^{*-}_s\tau^+\nu_\tau)$                &$\cdots$                    &$1.20\pm0.07$\\\hline
\end{tabular}\label{Tab:BrB2DDVlv}}
%
\renewcommand\arraystretch{1.2}
\tabcolsep 0.1in
\centering
\caption{The experimental data and the SU(3) flavor symmetry predictions of $\mathcal{B}(B\to D_0/D_2^*\ell^+\nu_\ell)$   within $2\sigma$ errors. $^{2q(4q)}$denote the two(four) quark state predictions.  $\mathcal{B}(B\to D_0/D^*_2\ell'^+\nu_{\ell'})$ is in unit of $10^{-3}$, and $\mathcal{B}(B\to D_0/D^*_2\tau^+\nu_\tau)$ is in unit of $10^{-4}$.
}\vspace{0.1cm}
{\footnotesize
\begin{tabular}{lcc|lccc}  \hline\hline
Branching ratios                                   & Exp. data  \cite{Belle-II:2022evt}            & Our predictions                               & Branching ratios                                                & Exp. data  \cite{Belle-II:2022evt}                                   & Our predictions        \\\hline
$\mathcal{B}(B^+\to\overline{D}^0_0\ell'^+\nu_{\ell'})$     &$4.2\pm1.6$                         &$3.98\pm1.30$                                  &$\mathcal{B}(B^+\to\overline{D}^{*0}_2\ell'^+\nu_{\ell'})$        &$2.9\pm0.6$ &$3.20\pm0.30$            \\
$\mathcal{B}(B^0\to D^-_0\ell'^+\nu_{\ell'})$               &$3.9\pm1.4$                         &$3.71\pm1.21$                                  &$\mathcal{B}(B^0\to D^{*-}_2\ell'^+\nu_{\ell'})$                  &$2.7\pm0.6$ &$2.99\pm0.29$             \\
$\mathcal{B}(B^0_s\to D^-_{s0}\ell'^+\nu_{\ell'})$            &$\cdots$                            &$4.43\pm1.54^{2q}$,~ $8.84\pm3.08^{4q}$      &$\mathcal{B}(B^0_s\to D^{*-}_{s2}\ell'^+\nu_{\ell'})$               &$\cdots$                                       &$2.72\pm0.27$            \\
$\mathcal{B}(B^+\to\overline{D}^0_0\tau^+\nu_\tau)$         &$\cdots$                            &$5.23\pm1.85$                                  &$\mathcal{B}(B^+\to\overline{D}^{*0}_2\tau^+\nu_\tau)$            &$\cdots$                                       &$2.15\pm0.47$             \\
$\mathcal{B}(B^0\to D^-_0\tau^+\nu_\tau)$                   &$\cdots$                            &$4.86\pm1.72$                                  &$\mathcal{B}(B^0\to D^{*-}_2\tau^+\nu_\tau)$                      &$\cdots$                                       &$1.97\pm0.43$              \\
$\mathcal{B}(B^0_s\to D^-_{s0}\tau^+\nu_\tau)$                &$\cdots$                            &$6.75\pm2.40^{2q}$,~ $13.50 \pm4.80^{4q}$            &$\mathcal{B}(B^0_s\to D^{*-}_{s2}\tau^+\nu_\tau)$                   &$\cdots$                                       &$1.73\pm0.38$              \\\hline
\end{tabular}\label{Tab:BrB2D0D2lv}}
\end{table}

For the $B\to D^* \ell^+\nu_\ell$ decays, there are also five measured modes, and they are listed in the fifth column of Tab. \ref{Tab:BrB2DDVlv}.  $\mathcal{B}(B^+\to \overline{D}^{*0}\tau^+\nu_\tau)$ and $\mathcal{B}(B^0\to D^{*-}\tau^+\nu_\tau)$ are not used to constrain the only  free parameter $V^{BD^*}(0)$.
We obtain  that $V^{BD^*}(0)=0.65\pm0.05$ from three experimental data of $\mathcal{B}(B_{(s)}\to D^* \ell'^+\nu_{\ell'})$, which is smaller than $0.76$ given in Ref. \cite{Totten:2000ab}.  Then one can predict other branching ratios of the  $B\to D^{*}\ell^+\nu_{\ell}$   decays, which  are listed in the last column of Tab. \ref{Tab:BrB2DDVlv}.
One can see that  our prediction and experimental data of $\mathcal{B}(B^+\to \overline{D}^{*0}\tau^+\nu_\tau)$ intersect within $2\sigma$ error ranges, nevertheless, our prediction  of  $\mathcal{B}(B^0\to D^{*-}\tau^+\nu_\tau)$  is slightly smaller than its data, and they will agree  within $3\sigma$ error ranges.

For the $B\to D_0 \ell^+\nu_\ell$ decays, only two decay modes have been measured, and they are listed in the second column of Tab. \ref{Tab:BrB2D0D2lv}, which are used to constrain the  parameter $f^{BD_0}_1(0)$. Our constrained $f^{BD_0}_1(0)=0.38\pm0.09$, which is obviously  larger than $0.27\pm0.03$ given in Ref. \cite{Verma:2011yw}.   Our  branching ratio predictions  of the   $B\to D_0\ell^+\nu_\ell$  decays   are listed in the third column of Tab. \ref{Tab:BrB2D0D2lv}. The branching ratio predictions of $B^0_s\to D^-_{s0}\ell^+\nu_{\ell}$  are  different between the two quark state  and the four quark state, the  prediction with four quark state are  2 times  of one with the two quark state.  In the later analysis of $B\to DP\ell\nu_\ell$ with $D_0$ resonances, the results of the two quark state will be used.

For the $B\to D^*_2 \ell^+\nu_\ell$ decays, only $\mathcal{B}(B^+\to\overline{D}^{*0}_2\ell'^+\nu_{\ell'})$ and $\mathcal{B}(B^0\to D^{*-}_2\ell'^+\nu_{\ell'})$   have been measured, and they are listed in the fifth column of Tab. \ref{Tab:BrB2D0D2lv}. We obtain   $A_1^{BD^*_2}(0)=0.46\pm0.06$ from two measured branching ratios,  which are consistent with $0.63^{+0.11}_{-0.12}$  within $2\sigma$ errors given in Ref. \cite{Chen:2021ywv}.  The  branching ratio predictions  of the  $B\to D^*_2\ell^+\nu_\ell$  decays are listed in the last column of Tab. \ref{Tab:BrB2D0D2lv}.
Decays $B^0_s\to D^{*-}_{s2}\ell^+\nu_\ell$ have been calculated by the QCD sum rule approach for different scale parameter $\mu=2/3/4$ GeV \cite{Azizi:2014nta}, for examples,
$\mathcal{B}(B^0_s\to D^{*-}_{s2}\ell'^+\nu_{\ell'})=(3.07\pm1.40)\times10^{-3}$ and
$\mathcal{B}(B^0_s\to D^{*-}_{s2}\tau^+\nu_\tau)=(1.03\pm0.61)\times10^{-3}$ for $\mu=4$ GeV.
The predictions of $\mathcal{B}(B^0_s\to D^{*-}_{s2}\ell'^+\nu_{\ell'})$ in Ref. \cite{Azizi:2014nta} are consistent with ours, nevertheless,  the predictions of $\mathcal{B}(B^0_s\to D^{*-}_{s2}\tau^+\nu_{\tau})$ in Ref. \cite{Azizi:2014nta}  are smaller than ours.

Until now, most of the SU(3) flavor symmetry predictions of the  $B\to D_J\ell^+\nu_\ell$
 decays are quite coincident with their experimental data within $2\sigma$ errors.
The SU(3) flavor breaking effects  mainly come from different masses of $u$, $d$, and $s$ quarks.  Since  $m_{u,d}$ are much smaller than $m_s$, the SU(3)
breaking effects due to a non-zero $m_s$ dominate the SU(3) breaking effects \cite{He:2014xha}.
If considering the SU(3) flavor breaking effects due to  a non-zero $m_s$,
the  non-perturbative coefficients of the $B^0_s\to D^-_s/D^{*-}_{s}/D^-_{s0}/D^{*-}_{s2}\ell^+\nu_{\ell}$ decays are different from those of the  $B^+\to \overline{D}^0/\overline{D}^{*0}/\overline{D}^0_0/\overline{D}^{*0}_2\ell^+\nu_{\ell}$ and $B^0\to D^-/D^{*-}/D^-_0/D^{*-}_2\ell^+\nu_{\ell}$ decays.
As given in Tab. \ref{Tab:BrB2DDVlv}, decays $B^0_s\to D^-_{s}\mu^+\nu_{\mu}$ and $B^0_s\to D^{*-}_{s}\mu^+\nu_{\mu}$  have been measured,
so one can estimate the  SU(3) flavor breaking effects due to  a non-zero $m_s$ in the $B\to D/D^*\ell^+\nu_\ell$ decays.
Comparing our SU(3) flavor symmetry predictions and their experimental measurements of $B^0_s\to D^-_s/D^{*-}_{s}\mu^+\nu_{\mu}$ decays within $2\sigma$ errors, one can find the SU(3) breaking contributions to $\mathcal{B}(B^0_s\to D^-_s\ell^+\nu_{\ell})$  and $\mathcal{B}(B^0_s\to D^{*-}_{s}\ell^+\nu_{\ell})$ should be less than roughly 23\% and 20\% of their experimental central values, respectively.
After the $B^0_s\to D^{-}_{s0}/D^{*-}_{s2}\ell^+\nu_{\ell}$ decays are measured, one can estimate the  SU(3) flavor breaking effects in the $B\to D_0/D^{*}_2\ell^+\nu_\ell$ decays.

\subsection{Decays $D_J\to DP$ }
Non-leptonic two-body $D_J\to DP$ decays via strong or electromagnetic interaction are presented  in this section.
In terms of the SU(3) flavor symmetry,  the  decay amplitudes of the strong or electromagnetic $D_J\to DP$ decays can be parameterized as
\begin{eqnarray}
A(D_J\to DP)=a^{D_J}_{01}(D_J)_iP^i_jD^j+a^{D_J}_{02}(D_J)_iD^iP^j_j,
\end{eqnarray}
where $a^{D_J}_{01,02}$ are the non-perturbative coefficients, and all $D_J$ are two-quark states. $a^{D_J}_{02}$ are OZI suppressed and it will be ignored in later numerical analysis.
The decay amplitudes  for each $D^*_2\to DP$ decay   are  summarized in
Tab. \ref{Tab:DJ2DPAmp}. The decay amplitudes  for $D^*/D_0\to DP$ can be obtained by replacing $a^{D^*_2}_{01,02}$ listed in Tab. \ref{Tab:DJ2DPAmp} with $a^{D^*,D_0}_{01,02}$ only  if their phase spaces are  allowed.

Then the branching ratios of the $D_J\to DP$ decays  can be written as  \cite{Cheng:2020ipp}
\begin{eqnarray}
\mathcal{B}(D_0\to DP)&=&\frac{\tau_{D_0}p_c(m_{D_0},m_D,m_P)}{8\pi m^2_{D_0}}|A(D_0\to DP)|^2,\\
\mathcal{B}(D^*\to DP)&=&\frac{\tau_{D^*}p_c^3(m_{D^*},m_D,m_P)}{6\pi m^2_{D^*}}|A(D^*\to DP)|^2,\label{Eq:DV2DP}\\
\mathcal{B}(D_2^*\to DP)&=&\frac{\tau_{D_2^*}p_c^5(m_{D_2^*},m_D,m_P)}{60\pi m^2_{D_2^*}}|A(D_2^*\to DP)|^2,
\end{eqnarray}
where the center of mass  momentum $p_c(m_{D_J},m_D,m_P)\equiv\frac{\sqrt{\lambda(m_{D_J}^2,m_D^2,m_P^2)}}{2m_{D_J}}$.

Four decay modes of the $D^*\to D\pi$ decays have been measured, and the data within $2\sigma$ errors are \cite{ParticleDataGroup:2022pth}
\begin{eqnarray}
\mathcal{B}(D^{*0}\to D^0\pi^0)&=&(64.7\pm1.8)\%, ~~~~~~~~~~~~~~~\mathcal{B}(D^{*+}\to D^0\pi^+)=(67.7\pm1.0)\%,\nonumber\\
\mathcal{B}(D^{*+}\to D^+\pi^0)&=&(30.7\pm1.0)\%,~~~~~~~~~~~~~~~\mathcal{B}(D^{*+}_s\to D^+_s\pi^0)=(5.8\pm1.4)\%. \label{BrDVp2Dpi2}
\end{eqnarray}
Isospin violating decays $D^{*+}_s\to D^+_s\pi^0$ and later $D_{s0}^0\to D^0_s\pi^0$, which might decay via $\eta-\pi^0$ mixing \cite{Cho:1994zu},  are not considered in this work.
The experimental data of $\mathcal{B}(D^{*0}\to D^0\pi^0)$ and $\mathcal{B}(D^{*+}\to D^0\pi^+,D^+\pi^0)$ will be used to obtain $\mathcal{B}(B\to D\pi\ell\nu_\ell)$ with the  $D^*$  resonances.

In the $D_0\to DP$ decays, only $D_0\to D\pi$ decays have been seen but have no data. Since $D_0\to D\eta$, $D_0\to D_sK$  and $D_{s0}\to DK$ are not allowed by the phase spaces, we assume $\mathcal{B}(D_0\to D\pi)=1$ to obtain four branching ratios of the $D_0^0\to D^0\pi^0,D^+\pi^-$ and $D_0^+\to D^+\pi^0,D^0\pi^+$ decays. And they are
\begin{eqnarray}
\mathcal{B}(D^{0}_0\to D^0\pi^0)&=&(33.62\pm0.03)\%,~~~~~~~~~~~~~~~\mathcal{B}(D^{0}_0\to D^+\pi^-)=(66.38\pm0.03)\%,\nonumber\\
\mathcal{B}(D^{+}_0\to D^+\pi^0)&=&(33.18\pm0.01)\%,~~~~~~~~~~~~~~~\mathcal{B}(D^{+}_0\to D^0\pi^+)=(66.82\pm0.01)\%.\label{BrD0p2Dpi2}
\end{eqnarray}
The branching ratios of $D_0\to D\pi$ decays are the same as those in Eq. (\ref{BrD0p2Dpi2})  when considering $D_0$  as the four quark state.

For the  $D^*_2\to DP$ decays, since there is no experimental data of the branching ratios, we can not constrain $a^{D^*_2}_{01}$ directly. Nevertheless, $\frac{\mathcal{B}(D^*_2\to D\pi)}{\mathcal{B}(D^*_2\to D^*\pi)}=1.52\pm0.14$ within $2\sigma$  errors are measured  \cite{ParticleDataGroup:2022pth}. The SU(3) relation of the decay amplitudes of  the $D^*_2\to D^*P$ decays are given in Ref. \cite{Wan:2024ssl}. Using $\frac{\mathcal{B}(D^*_2\to D\pi)}{\mathcal{B}(D^*_2\to D^*\pi)}=1.52\pm0.14$  and  assuming $\mathcal{B}(D^{*0}_2\to D\pi)+\mathcal{B}(D^{*0}_2\to D^*\pi)\leq1$,
$\mathcal{B}(D^{*+}_2\to D\pi)+\mathcal{B}(D^{*+}_2\to D^*\pi)\leq1$ and $\mathcal{B}(D^{*+}_{s2}\to DK)+\mathcal{B}(D^{*+}_{s2}\to D^*K)\leq1$,  one  can  constrain  the non-perturbative coefficients $a^{D^*_2}_{01}$, and we obtain that   $|a^{D^*_2}_{01}|=25.14\pm1.47$.
 Then the branching ratios of the $D^*_2\to DP$  decays can be predicted,   which are given in the second column of Tab. \ref{Tab:BrD22DP}.
In addition,  their decay width predictions  and previous width predictions are also given in the third and fourth  columns of Tab. \ref{Tab:BrD22DP}, respectively.  Our width predictions are about 1 time larger than ones in Ref. \cite{Wang:2014oca}, nevertheless, they are very close to ones in Ref. \cite{Zhang:2016dom}.

\begin{table}[t]
\renewcommand\arraystretch{1.45}
\tabcolsep 0.05in
\centering
\caption{The decay amplitudes for the $D^*_2\to DP$ decays under the  SU(3) flavor symmetry.
}\vspace{0.1cm}
{\footnotesize
\begin{tabular}{lc|lc}  \hline\hline
~~~~~Decay modes                               & Coupling vertexs                                                                                               &  ~~~~~Decay modes  & Coupling vertexs \\\hline
$D^{*0}_2\to D^0\pi^0$             &$a^{D^*_2}_{01}/\sqrt{2}$                                                                                          &$D^{*+}_2\to D^+\pi^0$                     &$-a^{D^*_2}_{01}/\sqrt{2}$   \\
$D^{*0}_2\to D^0\eta$              &$a^{D^*_2}_{01}\Big(\frac{cos\theta_P}{\sqrt{6}}-\frac{sin\theta_P}{\sqrt{3}}\Big)-\sqrt{3}a^{D^*_2}_{02}sin\theta_P$ &$D^{*+}_2\to D^+\eta$                      &$a^{D^*_2}_{01}\Big(\frac{cos\theta_P}{\sqrt{6}}-\frac{sin\theta_P}{\sqrt{3}}\Big)-\sqrt{3}a^{D^*_2}_{02}sin\theta_P$  \\
$D^{*0}_2\to D^+\pi^-$             &$a^{D^*_2}_{01} $                                                                                                  &$D^{*+}_2\to D^0\pi^+$                     &$a^{D^*_2}_{01} $  \\
$D^{*0}_2\to D^+_s K^-$            &$a^{D^*_2}_{01} $                                                                                                  &$D^{*+}_2\to D^{+}_s \overline{K}^0$       &$a^{D^*_2}_{01} $    \\
$D^{*+}_{s2}\to D^0K^+$            &$a^{D^*_2}_{01}$\\
$D^{*+}_{s2}\to D^+K^0$            &$a^{D^*_2}_{01} $  \\
$D^{*+}_{s2}\to D^+_s\eta$         &$-a^{D^*_2}_{01}\Big(\frac{2cos\theta_P}{\sqrt{6}}+\frac{sin\theta_P}{\sqrt{3}}\Big)-\sqrt{3}a^{D^*_2}_{02}sin\theta_P$\\ \hline
\end{tabular}\label{Tab:DJ2DPAmp}}
\renewcommand\arraystretch{1.45}
\tabcolsep 0.15in
\centering
\caption{ The branching ratio predictions of the $D^*_2\to DP$ decays within $2\sigma$ errors.  }\vspace{0.1cm}
{\footnotesize
\begin{tabular}{lccc}  \hline\hline
Decay modes  & Branching ratios $(\times10^{-2})$ & Decay widthes (MeV) &  Decay widthes from  (MeV)   \\\hline
$D^{*0}_2\to D^0\pi^0$                &$20.18 \pm1.77 $                   &$9.59 \pm 1.06$                    & $4.14^{+1.82}_{-1.57}$ \cite{Wang:2014oca}, $12.0$ \cite{Zhang:2016dom} \\
$D^{*0}_2\to D^0\eta$                 &$0.13  \pm0.03 $                   &$0.06 \pm0.02 $                    & $\cdots$\\
$D^{*0}_2\to D^+\pi^-$                &$38.51 \pm3.41 $                   &$18.30 \pm2.03 $                   & $7.91^{+3.49}_{-3.00} $ \cite{Wang:2014oca}, $22.8$ \cite{Zhang:2016dom}\\
$D^{*0}_2\to D^+_s K^-$               &$ [1.82\times10^{-15},6.56\times10^{-6}] $     &$ [8.77\times10^{-14},3.00\times10^{-6}] $    & $\cdots$\\
$D^{*+}_2\to D^+\pi^0$                &$19.40 \pm1.63 $                   &$9.33 \pm0.92 $                    & $\cdots$\\
$D^{*+}_2\to D^+\eta$                 &$0.11 \pm0.04 $                    &$ (5.45 \pm2.15)\times10^{-2}$     & $\cdots$\\
$D^{*+}_2\to D^0\pi^+$                &$39.88 \pm3.33 $                   &$19.18 \pm1.90 $                   & $\cdots$\\
$D^{*+}_2\to D^{+}_s \overline{K}^0$  &$ [1.26\times10^{-14},1.80\times10^{-6}] $     &$ [6.01\times10^{-13},8.28\times10^{-5}] $   & $\cdots$\\
$D^{*+}_{s2}\to D^0K^+$               &$42.32 \pm5.20 $                   &$7.52 \pm0.95 $                   & $3.35^{+1.48}_{-1.27}$ \cite{Wang:2014oca}, $9.49$ \cite{Zhang:2016dom} \\
$D^{*+}_{s2}\to D^+K^0$               &$38.12 \pm4.72 $                   &$6.79 \pm0.86 $                   & $3.04^{+1.34}_{-1.15}$ \cite{Wang:2014oca}, $8.61$ \cite{Zhang:2016dom} \\
$D^{*+}_{s2}\to D^+_s\eta$            &$0.58 \pm0.19 $                    &$(10.23 \pm3.43)\times10^{-2} $    & $\cdots$\\ \hline
\end{tabular}\label{Tab:BrD22DP}}
\end{table}

\subsection{ Numerical results of the resonant $B\to D P\ell^+\nu_\ell$ decays}
In terms of  $\mathcal{B}(B \rightarrow D_J \ell^+\nu_\ell)$ given in Tabs. \ref{Tab:BrB2D0D2lv}-\ref{Tab:BrB2DDVlv}  and $\mathcal{B}(D_J \rightarrow D P)$ given in Eqs. (\ref{BrDVp2Dpi2}-\ref{BrD0p2Dpi2}) and Tab. \ref{Tab:BrD22DP},  after  considering the further experimental bounds of the resonant $B\to D \pi\ell^+\nu_\ell$ decays given in Eqs. (\ref{EBrBu2Dpilv0}-\ref{EBrBu2Dpilv2}) and (\ref{EBrBd2Dpilv0}-\ref{EBrBd2Dpilv2}), one can obtain the branching ratio predictions of the resonant $B\to D P\ell^+\nu_\ell$ decays, and they are listed in the third, fourth and fifth columns of Tab. \ref{Tab:BrD2DPlv} for the $D^*$, $D_0$ and $D_{(s)2}^*$ resonances, respectively.  Corresponding experimental data with $2\sigma$ errors are also listed in Tab. \ref{Tab:BrD2DPlv} for   the convenience of comparison.  Note that, since the vector resonances are also considered in this work,   $\mathcal{B}(B^+\to D^{-}\pi^+\ell'^+\nu_{\ell'})_{T}$ and $\mathcal{B}(B^0\to D^{0}\pi^-\ell'^+\nu_{\ell'})_{T}$ in Eq. (\ref{EBrBu2DpilvA}) and Eq. (\ref{EBrBd2DpilvA}), which only conclude $D_0$ and $D^*_2$ resonances,  are not  used for our results.
Many  resonant branching ratios in Tab. \ref{Tab:BrD2DPlv} are predicted for the first time.

One can see that the the vector meson $D^*$ resonances  give the  dominant contributions in the $B^+\to \overline{D}^0\pi^0\ell^+\nu_{\ell}$, $B^0\to \overline{D}^0\pi^-\ell^+\nu_{\ell}$ and $B^0\to D^-\pi^0\ell^+\nu_{\ell}$ decays, largely because of its proximity to the $D\pi$ threshold. Please note that  decay amplitude of the  $B^+\to D^-\pi^+\ell^+\nu_\ell$ decays  is larger than ones of the $B^+\to D^0\pi^0\ell^+\nu_\ell$ decays
by factor $\sqrt{2}$, nevertheless,  the latter branching ratios are   much larger than the  former ones, since  the most dominant resonance $D^{*0}$ cannot decay into $D^-\pi^+$ on its mass-shell \cite{ParticleDataGroup:2022pth}. In previous studies, $\mathcal{B}(B^+\to \overline{D}^0\pi^0\ell'^+\nu_{\ell'})_{D^{*0}}=34.9\times10^{-3}$ \cite{Cheng:1993ah}, $\mathcal{B}(B^0\to D^-\pi^0\ell'^+\nu_{\ell'})_{D^{*-}}=16.7\times10^{-3}$ \cite{Cheng:1993ah},
$\mathcal{B}(B^0\to D^-\pi^0\ell'^+\nu_{\ell'})_{D^{*-}}=14.0\times10^{-3}$ \cite{Kim:2017dfr},
$\mathcal{B}(B^0\to D^-\pi^0\tau^+\nu_{\tau})_{D^{*-}}=3.53\times10^{-3}$ \cite{Kim:2017dfr},
after considering the error, our corresponding results are consistent with them.

As for the scalar meson $D_0$ resonances and the tensor meson $D^*_2$ resonances, the experimental upper limit of $\mathcal{B}(B^+\to D^-\pi^+\ell'^+\nu_{\ell'})_{D^0_0}$ gives  further constraint on the $\mathcal{B}(B\to D\pi\ell'^+\nu_{\ell'})_{D_0}$ predictions, and  the experimental lower limit of $\mathcal{B}(B^+\to D^-\pi^+\ell'^+\nu_{\ell'})_{D^{*0}_2}$ gives  further constraint on the $\mathcal{B}(B\to D\pi\ell'^+\nu_{\ell'})_{D^*_2}$ predictions.
Our predictions for $\mathcal{B}(B^0\to \overline{D}^0\pi^-\ell'^+\nu_{\ell'})_{D^-_0,D^{*-}_2}$ are more precise than their experimental measurements.
The contributions of the  $D_0$ and $D^{*}_2$ resonances are  in the same order of magnitude in the $B^+\to D^-\pi^+\ell'^+\nu_{\ell'}$ and $B^0\to \overline{D}^0\pi^-\ell'^+\nu_{\ell'}$ decays. But the contributions of the  $D_0$  resonances are larger than ones of the $D^{*}_2$  resonances in the $B^+\to \overline{D}^0\pi^0\ell^+\nu_{\ell}$, $B^0\to D^-\pi^0\ell^+\nu_{\ell}$, $B^+\to D^-\pi^+\tau^+\nu_{\tau}$ and $B^0\to \overline{D}^0\pi^-\tau^+\nu_{\tau}$ decays.

The Belle II experiment has reported the branching ratios of the $B\to D\eta\ell'^+\nu_{\ell'}$ decays with quite large errors,  $\mathcal{B}(B^0\to D^{-}\eta\ell'^+\nu_{\ell'})_{T}= (4.0\pm4.0)\times10^{-3}$ and
$\mathcal{B}(B^+\to D^{0}\eta\ell'^+\nu_{\ell'})_{T}= (4.0\pm4.0)\times10^{-3}$ \cite{Belle-II:2022evt}, which are not used for our predictions.  From our predictions, one can see that the non-resonant branching ratios are dominant in the $B\to D\eta\ell'^+\nu_{\ell'}$ decays. And our predictions of $\mathcal{B}(B\to D\eta\ell'^+\nu_{\ell'})_N$ lie in the range of experimental data with $1\sigma$ error.

In addition,  the interference terms between the non-resonant, the vector resonant, the scalar resonant and  the  tensor resonant contributions  exist,
and they might not be ignored if  more than one kind of contributions  are important in the decays, and they will be studied in our succeeding work.

All present experimental data of $\mathcal{B}(B\to D P\ell^+\nu_\ell)$  may be explained by the SU(3) flavor symmetry approach.
The SU(3) flavor breaking effects in the $B\to D P\ell^+\nu_\ell$  decays are in the similar  to ones in the $B\to D^* P\ell^+\nu_\ell$  decays.
As given in Tab. \ref{Tab:BrD2DPlv},  the dominate SU(3) flavor breaking effects might appear  in the non-resonant and the charmed tensor resonant $B\to D P\ell^+\nu_\ell$ decays.
Nevertheless, there is only the  data for $\mathcal{B}(B_{u,d}\to D \pi\ell^+\nu_\ell)_{D^*_2}$, and there is not any data for $\mathcal{B}(B_{s}\to D K\ell^+\nu_\ell)_{D^*_{s2}}$.  Or  there is only the  data of $\mathcal{B}(B^+\to D_s^- K^+ \ell^+\nu_\ell)_N$, and there is no   data of $\mathcal{B}(B_{u,d,s}\to D K\ell^+\nu_\ell)_N$.
Therefore,  we can not directly judge how  large  the possible  SU(3) breaking effects  are in the $B\to D P\ell^+\nu_\ell$ decays.

Although the widths of all resonances are narrow, following Refs. \cite{Cheng:1993ah,LeYaouanc:2018zyo}, the  width effects of $D^*$ mesons are analyzed. After considering the width effects of $D^*$ mesons,
the decay branching ratios of $B\to D \pi\ell^+\nu_\ell$ are \cite{Cheng:1993ah,Tsai:2021ota}
{\small
\begin{eqnarray}
\mathcal{B}(B\to D^* \pi\ell^+\nu_\ell)_{D^*}=\frac{1}{\pi}\int_{(m_{D^*}-n\Gamma_{D^*})^2}^{(m_{D^*}+n\Gamma_{D^*})^2} dt_V \int_{m^2_{\ell}}^{(m_B-\sqrt{t_V})^2} dq^2\frac{\sqrt{t_V}d\mathcal{B}(B\to D^*(\lambda)\ell^+\nu_\ell,t_V)/dq^2~\mathcal{B}(D^*(\lambda)\to D\pi,t_V)\Gamma_{D^*}}{(t_V-m_{D^*}^2)^2+m^2_{D^*}\Gamma^2_{D^*}},\label{Eq:RDA4}
\end{eqnarray}}
where $d\mathcal{B}(B\to D^*(\lambda)\ell^+\nu_\ell,t_V)/dq^2$ and $\mathcal{B}(D^*(\lambda)\to D\pi,t_V)$ are obtained from  to Eq. (\ref{Eq:dbdq2}) and Eq. (\ref{Eq:DV2DP}) by replacing $m_{D^*} \to \sqrt{t_V}$, respectively.
There are two non-perturbative coefficients  $V^{BD^*}(0)$  in $d\mathcal{B}(B\to D^*(\lambda)\ell^+\nu_\ell,t_V)/dq^2$ and $a_{01}^{D^*}$   in $\mathcal{B}(D^*(\lambda)\to D\pi,t_V)$.
$V^{BD^*}(0)=0.65\pm0.05$ from the data of $\mathcal{B}(B\to D^*\ell^+\nu_\ell)$ listed in Tab. \ref{Tab:BrB2DDVlv}, and
$a_{01}^{D^*}=8.42\pm0.38$ from the data of $\mathcal{B}(D^{*+}\to D\pi)$ given in Eq. (\ref{BrDVp2Dpi2}).
$\Gamma_{D^{*0}}=(56.00\pm5.74)~KeV$, which is obtained via the experimental data of $\mathcal{B}(D^{*0}\to D^0\pi^0)$ in Eq. (\ref{BrDVp2Dpi2}) and the SU(3) flavor symmetry in $D^*\to D\pi$.
Following Refs. \cite{Cheng:1993ah}, choosing $n=3$, the results of  $\mathcal{B}(B\to D \pi\ell^+\nu_\ell)_{D^*}$ are obtained, and they are listed in Tab. \ref{Tab:BrD2DPlv}  by denoting $\dagger$. One can see that the results obtained by considering the $D^*$ width effects are slightly smaller than ones obtained by the narrow width  approximation.

\section{Summary}

The semileptonic $B\to DP\ell^+\nu_\ell$  decays with  the non-resonances, the vector resonances, the scalar resonances and  the tensor resonances have been explored  in terms of  the SU(3) flavor symmetry based on the relevant experimental data.
The amplitude relations of the non-resonant $B\to DP\ell^+\nu_\ell$ decays, the semileptonic $B\to D/D^*/D_0/D^*_2\ell^+\nu_\ell$ decays and the non-leptonic $D^*/D_0/D^*_2\to DP$ decays  have been obtained,  and then the resonant branching ratios have been obtained by the narrow width approximation after considering   the resonant  experimental data of the $B\to DP\ell'^+\nu_{\ell'}$ decays.  Our main results can be summarized as follows.

For the non-resonant $B\to DP\ell^+\nu_{\ell}$ decays, the central values of
$\mathcal{B}(B^+\to D^-\pi^+\ell'^+\nu_{\ell'})_N$, $\mathcal{B}(B^+\to \overline{D}^0\pi^0\ell'^+\nu_{\ell'})_N$, $\mathcal{B}(B^+\to \overline{D}^0\eta\ell'^+\nu_{\ell'})_N$, $\mathcal{B}(B^0\to \overline{D}^0\pi^-\ell'^+\nu_{\ell'})_N$
$\mathcal{B}(B^0\to D^-_sK^0\ell'^+\nu_{\ell'})_N$, $\mathcal{B}(B^0\to D^-\pi^0\ell'^+\nu_{\ell'})_N$, $\mathcal{B}(B^0\to D^-\eta\ell'^+\nu_{\ell'})_N$, $\mathcal{B}(B^0_s\to \overline{D}^0K^-\ell'^+\nu_{\ell'})_N$
$\mathcal{B}(B^0_s\to D^-\overline{K}^0\ell'^+\nu_{\ell'})_N$ and  $\mathcal{B}(B^0_s\to D_s^-\eta\ell'^+\nu_{\ell'})_N$   are on the orders
of $10^{-4}$, which could be measured by the LHCb and Belle II experiments. Other non-resonant decays
are strongly suppressed by the narrow phase spaces.

For the charmed vector resonant  $B\to DP\ell^+\nu_{\ell}$ decays,  they give the  dominant contributions in the $B^+\to \overline{D}^0\pi^0\ell^+\nu_{\ell}$, $B^0\to \overline{D}^0\pi^-\ell^+\nu_{\ell}$ and $B^0\to D^-\pi^0\ell^+\nu_{\ell}$ decays.  Nevertheless, since the  resonance $D^{*0}$ can not decay into $D^-\pi^+$, the total branching ratios of the  $B^+\to D^-\pi^+\ell^+\nu_\ell$ decays are    much smaller than ones of the $B^+\to \overline{D}^0\pi^0\ell^+\nu_{\ell}$, $B^0\to \overline{D}^0\pi^-\ell^+\nu_{\ell}$ and $B^0\to D^-\pi^0\ell^+\nu_{\ell}$ decays.

As for the  charmed scalar and tensor resonant  $B\to DP\ell^+\nu_{\ell}$ decays,   our predictions of $\mathcal{B}(B^0\to \overline{D}^0\pi^-\ell'^+\nu_{\ell'})_{D^-_0,D^{*-}_2}$ are more precise than their experimental measurements.
The contributions of the  $D_0$ and $D^{*}_2$ resonances are  in the same order of magnitude in the $B^+\to D^-\pi^+\ell'^+\nu_{\ell'}$ and $B^0\to \overline{D}^0\pi^-\ell'^+\nu_{\ell'}$ decays. But the contributions of the  $D_0$  resonances are larger than ones of $D^{*}_2$  resonances in the $B^+\to \overline{D}^0\pi^0\ell^+\nu_{\ell}$, $B^0\to D^-\pi^0\ell^+\nu_{\ell}$, $B^+\to D^-\pi^+\tau^+\nu_{\tau}$ and $B^0\to \overline{D}^0\pi^-\tau^+\nu_{\tau}$ decays.

Although the SU(3) flavor symmetry only can give the approximate predictions, they are still very useful for understanding these decays.
Until now, our predictions of the $B\to DP\ell^+\nu_{\ell}$ decays  are  quite coincident with present experimental data, and our predictions   could be   tested in future experiments, such as LHCb  and Belle II.

\section*{ACKNOWLEDGEMENTS}
The work was supported by the National Natural Science Foundation of China (No. 12175088 and No. 12365014).

\section*{References}

\end{document}